\documentclass[review]{elsarticle}
\usepackage{graphics}
\usepackage{graphicx}
\usepackage{float}
\usepackage{bm}
\usepackage[nomarkers,notablist,nofiglist]{endfloat}
\usepackage{subfigure}
\usepackage{lineno}
\modulolinenumbers[5]
\usepackage{siunitx}         
\graphicspath{ {./Figures/} } 
\usepackage{soul}
\usepackage{miller}
\usepackage{amsmath}
\usepackage{geometry} 
\usepackage{color} 
\usepackage{outlines}
\usepackage{upgreek}
\usepackage{multirow}

\usepackage{booktabs}

\journal{arXiv}

\begin{document}
\begin{frontmatter}

\title{Processing-dependent Chemical Ordering in a Metallic Alloy Characterized via Non-destructive Bragg Coherent Diffraction Imaging}

\author[1]{Nathaniel Warren}
\author[1]{Chloe Skidmore}
\author[2]{Katherine J. Harmon}
\author[3]{Wonsuk Cha}
\author[1]{Jon-Paul Maria}
\author[2]{Stephan O. Hruszkewycz}
\author[1]{Darren C. Pagan\corref{mycorrespondingauthor}}
 	\ead{dcp5303@psu.edu}
\address[1]{Materials Science and Engineering, Pennsylvania State University}
\address[2]{Materials Science Division, Argonne National Laboratory}
\address[3]{X-ray Science Division, Argonne National Laboratory}

\begin{abstract}
Of current importance for alloy design is controlling chemical ordering through processing routes to optimize an alloy’s mechanical properties for a desired application. However, characterization of chemical ordering remains an ongoing challenge, particularly when nondestructive characterization is needed. In this study, Bragg coherent diffraction imaging is used to reconstruct morphology and lattice displacement in model Cu$_3$Au nanocrystals that have undergone different heat treatments to produce variation in chemical ordering. The magnitudes and distributions of the scattering amplitudes (proportional to electron density) and lattice strains within these crystals are then analyzed to correlate them to the expected amount of chemical ordering present. Nanocrystals with increased amounts of ordering are found to generally have less extreme strains present and reduced strain distribution widths. In addition, statistical correlations are found between the spatial arrangement of scattering amplitude and lattice strains.

\end{abstract}

\end{frontmatter}

Chemical ordering is an important microstructural feature in many metallic alloys (along with other systems such as energy storage materials) which can contribute to mechanical properties \cite{cohen1962some}. The ordering can extend over long ranges to produce distinct phases or be limited to small (order nm) regions  where the local atomic arrangement preferentially follows a defined structure rather than that of a random solid solution. The latter is often referred to as short range order (SRO). The presence of ordered regions of limited size alters mechanical properties, such as strength, as there is an increased energy cost for dislocation propagation (crystallographic slip at larger length scales) through the ordered regions that is associated with altering the favorable ordered atomic arrangement \cite{rudman1956effect}. In other words, slip will be more difficult than in a random solution since it will require more energy to occur, and there is a corresponding increase in material strength. While the existence of SRO in metallic alloys is well-established, the phenomenon has recently grown in importance due to its prevalence in multi-principal element alloys (MPEAs) \cite{zhang2020short,han2024ubiquitous}. Control of SRO in MPEAs (and traditional alloy systems) is important for generating microstructures with superior properties \cite{wu2021short}.

Property optimization in metallic alloys first requires rigorous characterization to identify ordering-property correlations. Historically, the primary means to measure SRO is through diffuse X-ray (or neutron) scattering \cite{cowley1950approximate}. In these methods, preferential site occupancy produces enhanced broad (i.e., diffuse) scattering generally in the same locations as superlattice peaks of a fully-ordered structure which is then fit to an analytical model. However, analysis of these data has significant challenges, particularly deconvolution from other diffuse scattering, primarily thermal diffuse scattering, and an inability to spatially locate regions with SRO. More recently, pair distribution function (PDF) measurements have been utilized to quantify SRO \cite{proffen2002chemical,owen2017analysis}. Rather than diffuse scattering, SRO's effects on the entire scattered spectrum (both diffraction peaks and regions between peaks) are fit using forward projection models. Again, this analysis is susceptible to issues with model fitting. Crucially, diffuse scattering and PDF yield average structures and lack insights to SRO spatial heterogeneity throughout the diffraction volume. In contrast, electron diffraction is frequently employed because it allows for real-space observation of ordering over much smaller volumes with nanoscale resolution \cite{okamoto1971short,stobbs1978classification}. The ordered regions can even be directly observed \cite{sato2011order}, which has recently been extended to three dimensions using atomic electron tomography \cite{moniri2023three}. Similarly, atom probe tomography is also able to precisely reconstruct atomic positions in 3D \cite{marceau2015quantitative,he2024quantifying}. However, both electron and atom probe methods are inherently destructive.

To provide a complementary means to non-destructively characterize the 3D distribution of chemical ordering in crystalline materials, we demonstrate here the use of Bragg Coherent Diffraction Imaging (BCDI). BCDI is an X-ray characterization technique that utilizes coherent hard X-ray beams from synchrotron or X-ray free-electron laser sources to facilitate reconstruction of lattice displacement fields (and strains) within micro-scale crystallites at resolutions on the order 10 nm \cite{robinson2001reconstruction,pfeifer2006three,robinson2009coherent,vaxelaire2014new}. Here, BCDI is used to characterize the lattice inhomogeneity, particularly strain fields, within Cu$_3$Au nanocrystals that have undergone heat treatment to produce varying degrees of chemical ordering. Cu$_3$Au is a classic model material for SRO studies \cite{cowley1950x} where the relationship between ordering and strength has been established \cite{kear1964dislocation}. The changes in the strain magnitudes and the spatial distributions of strains are correlated to the extent of ordering that has developed.

To evaluate different microstructural characteristics using BCDI, three sets of Cu$_3$Au nanocrystals were produced with varied heat treatments to produce differing degrees of chemical ordering. First, films were deposited using magnetron co-sputtering with copper and gold targets on a (001) $\gamma$-Al$_2$O$_3$ substrate to a thickness of approximately 35 nm. Sputtering was performed for approximately one minute while flowing Ar gas at a pressure of 5 $\times$ 10$^{-3}$ torr. The deposition rates of Cu and Au were 30 nm/minute and 10 nm/minute, respectively, to achieve the desired composition of Cu$_3$Au. The Cu$_3$Au films were grown such that \{111\} lattice plane normals were preferentially aligned with the film normal. Nanocrystals were then formed by heating the thin films in a rapid annealer at 800 $^\circ$C for 30 minutes, which yielded alloy nanocrystals in a fully disordered solid solution. After this annealing, the nanocrystals ranged from 100 to 500 nm in size and retained the preferred alignment of \{111\} normals with the substrate normal. One substrate coated with nanocrystals was left in the fully disordered state, while others were heat treated further to induce different amounts of chemical order. One sample was heated at 403$^\circ$C for 12 hours to produce isolated regions of ordering (i.e., SRO) and a second was heated at 372$^\circ$C for 48 hours to try to maximize ordering. In total, these heat treatments produced samples with nanocrystals that contain increasing amounts of chemical ordering and are labeled `disordered', `semi-ordered', and `ordered' which are summarized in Table \ref{tab:ht_tab}.

\begin{table}
\centering
 \caption{Summary of the different heat treatments performed on the Cu$_3$Au nanocrystals to produced different ordering states.}
 \vspace{3mm}
\begin{tabular}{|c|c|}
\hline
Name & Heat Treatment Description\\ \hline
      Disordered & 800$ ^\circ$C for 0.5 hours\\
      Semi-ordered & 800$ ^\circ$C for 0.5 hours + 403$^\circ$C for 12 hours \\
      Ordered& 800$ ^\circ$C for 0.5 hours + 372$^\circ$C for 48 hours \\ \hline
\end{tabular}

    	 \label{tab:ht_tab}
\end{table}

\begin{figure}[h!]
    \centering
    \includegraphics[width = 0.8 \textwidth]{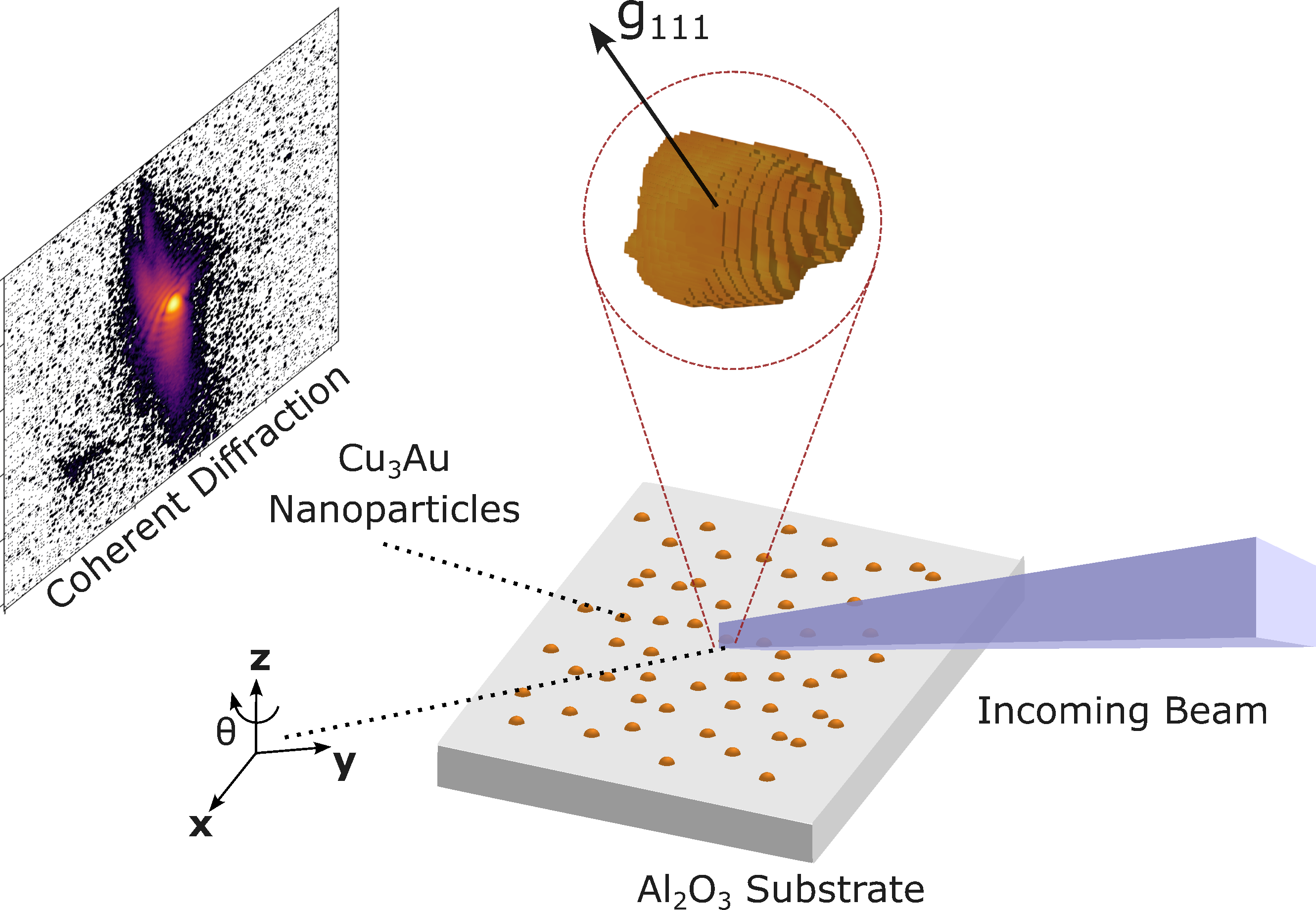}
    \caption{Schematic of the Bragg coherent diffraction experimental geometry used in this work. Cu$_3$Au nanocrystals on an Al$_2$O$_3$ substrate we illuminated by a coherent X-ray beam as diffracted intensity from a 111 diffraction peak were collected.}\label{fig:geometry}
\end{figure}

BCDI measurements were conducted at the 34-ID-C beamline of the Advanced Photon Source at Argonne National Laboratory. A schematic of the measurements is shown in Figure \ref{fig:geometry}. During measurements, specimens were illuminated by a focused 10 keV X-ray beam traveling along the $-\bm{y}$ direction. The beam was defined by a 25 $\mu$m $\times$ 50 $\mu$m slit that produced a 670 nm $\times$ 570 nm (horizontal $\times$ vertical) focused spot at the sample. The coherent diffraction data was collected on an area detector sitting 1 m away from the specimen. A Timepix pixelated area detector (Amsterdam Scientific Instruments) with 512 pixels $\times$ 512 pixels and a 55 $\mu$m pixel size was used. The detector was mounted on a moving arm and positioned to intersect a small angular portion of the Debye-Scherrer ring such that  Cu$_3$Au 111 diffraction peaks from individual nanocrystals could be captured on the detector. Nanocrystals for BCDI imaging were located by rastering the sample until diffracted intensity from a single nanocrystal was observed in the detector. Note that several nanocrystals with different lattice orientations were simultaneously illuminated in this configuration, but a single nanocrystal diffracted into the detector at a time. Once a nanocrystal was located, the samples were rocked by angle $\theta$ around $\bm{z}$ through the Bragg condition. The rocking consisted of 100 steps with 0.005$^\circ$ increments as coherent diffraction patterns were collected at each increment with the area detector to measure the full 3D distribution of intensity in reciprocal space. Exposure time for each pattern was 2.5 s, and the total scan took approximately 4 minutes. For each sample with different heat treatment, 3 nanocrystals were independently scanned. Preliminary BCDI reconstructions and screening of the nanocrystals were performed during data collection to select crystals approximately 300 nm in size and that maintained relatively equiaxed shapes during heat treatment.

The Python-based Phaser code, developed at Argonne National Laboratory and described in detail in \cite{maddali2020general,maddali2023concurrent}, was used to reconstruct the real-space (denoted by the real-space vector $\bm{r}$) 3D scattering field $\Psi(\bm{r})$ of the Cu$_3$Au nanocrystals from the diffracted intensity $I$ mapped to reciprocal space as a function of the scattering vector $\bm{q}$. The Phaser algorithm uses an iterative phase retrieval procedure to solve
\begin{equation}
I(\bm{q})=I_0 \left\Vert  \int \Psi(\bm{r}) e^{-i 2\pi \bm{q}\cdot \bm{r}} d\bm{r}  \right\Vert^2
\end{equation}
where $I_0$ is the incoming beam intensity. The scattering field $\Psi$ is a complex number described by an amplitude $A$ (i.e., the scattering power proportional to the electron density), and a phase $\phi$ associated with a displacement field $\bm{u}(\bm{r})$ projected along the scattering vector
\begin{equation}
\Psi(\bm{r})=A(\bm{r})e^{i\phi({\bm{r}})}=A(\bm{r})e^{-i 2\pi \bm{q}\cdot \bm{u}(\bm{r})} \quad .
\end{equation}
From the phase, a lattice strain field $\varepsilon_{hkl}$ for the probed set of lattice planes with normal $\bm{n_{hkl}}$ (or reciprocal lattice vector $\bm{g_{hkl}}$) can be calculated \cite{robinson2009coherent}
\begin{equation}
\varepsilon_{hkl}(\bm{r})=\bm{n_{hkl}}\cdot \bm{\varepsilon}(\bm{r})\cdot \bm{n_{hkl}}=\frac{\bm{n_{hkl}}}{\Vert\bm{g_{hkl}}\Vert} \cdot \nabla \phi(\bm{r}) 
\end{equation}
where $\bm{\varepsilon}$ is the lattice strain tensor and noting that during diffraction
\begin{equation}
\bm{q}=\bm{g_{hkl}}=\Vert\bm{g_{hkl}}\Vert\bm{n_{hkl}} \quad .
\end{equation}
The strain resolution for BCDI reconstructions has been reported to be approximately $2 \times 10^{-4}$ \cite{hofmann2018glancing,hofmann2020nanoscale}. We note here that the lattice strain is interpreted as a local deviation from an average lattice parameter that is due to chemical ordering and mechanical elastic strain. The spatial resolution of the reconstruction is anisotropic, dictated by the X-ray coherence lengths, X-ray energy, pixel size of the detector, rocking curve scan range and step intervals, and detector positioning. For this work, each voxel of the reconstruction has dimension of approximately 10 nm $\times$ 10 nm $\times$ 30 nm.

The Phaser algorithm begins the reconstruction by randomly instantiating the real space distribution of amplitude $A(\bm{r})$ and phase $\phi({\bm{r}})$, followed by a series of steps to reconstruct the scattering field. The primary reconstruction steps consist of a gradient-descent-driven update of the scattering field, where Fourier and inverse Fourier transforms are repeated until the difference between the intensity calculated from the Fourier transform of the real space object and the measured diffraction patterns is minimized (termed error reduction, ER)  \cite{robinson2009coherent}. A secondary reconstruction refinement is decreasing a particle shape function (or support) in order to reduce spurious solutions (termed shrink wrapping, SW). The reconstruction also intermittently performs perturbations of the reconstructions (termed hybrid-input output, HIO, and solvent flipping, SF \cite{fienup1982phase}) to attempt to prevent the reconstruction from becoming trapped in a local minimum during gradient descent. The optimized reconstruction algorithm (as determined by a reconstruction parameter sensitivity study) for the Cu$_3$Au nanocrystals consisted of 200 iterations of ER, 90 iterations of SF, 200 iterations of ER, 2000 iterations of HIO, 200 iterations of ER, SW (kernel size of 3 voxels and threshold of 0.05), 200 iterations of ER, 90 iterations of SF, 1000 iterations of ER.

\begin{figure}[h!]
    \centering
    \includegraphics[width = 1.0 \textwidth]{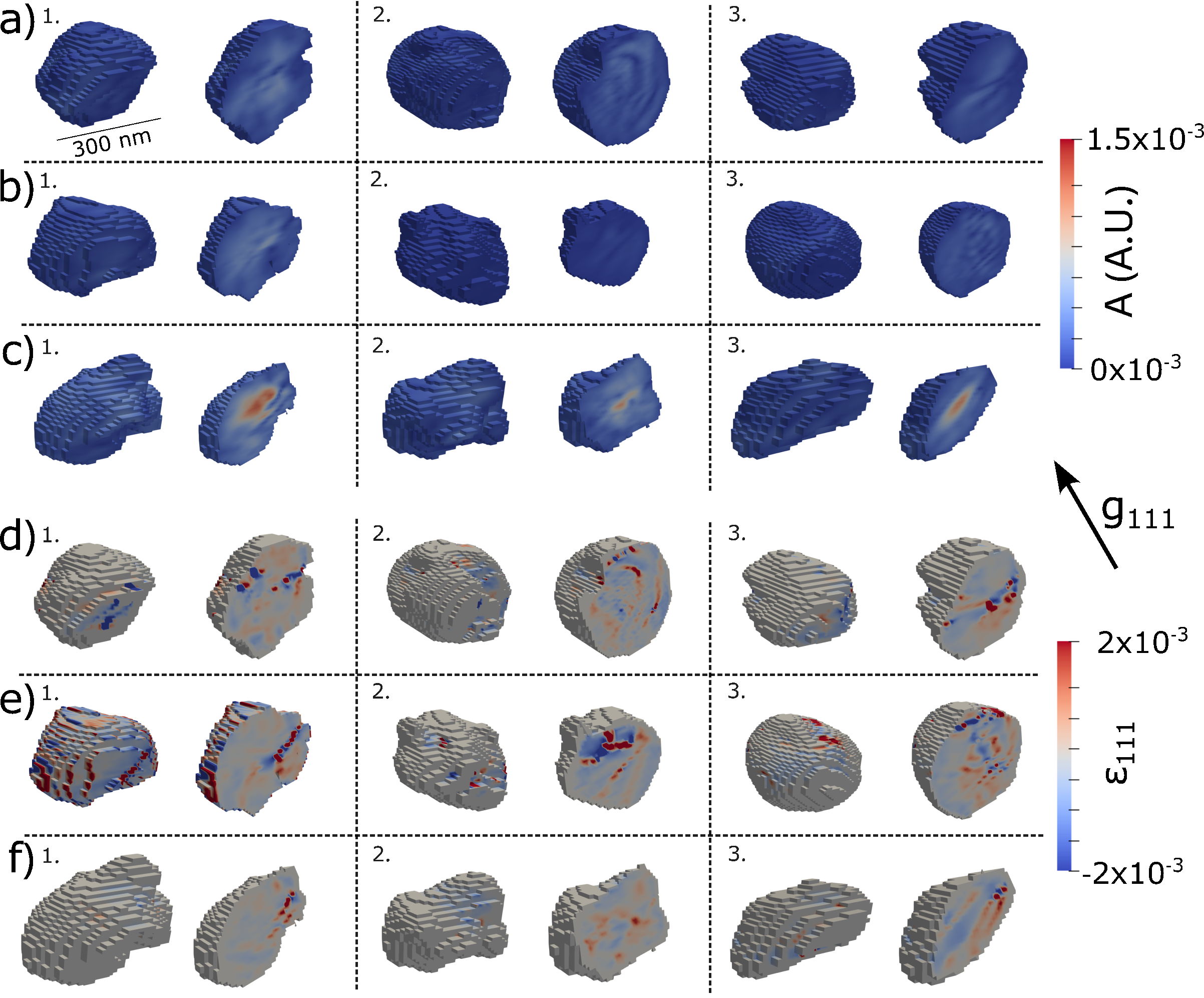}
    \caption{Reconstructed Cu$_3$Au scattering amplitude fields $A$ (both exterior and cross section) from the a) 3 disordered , b) 3 semi-ordered , and c) 3 ordered nanocrystals. Corresponding Cu$_3$Au 111 lattice strain fields $\varepsilon_{111}$ (both exterior and cross section) from the d) 3 disordered , e) 3 semi-ordered , and f) 3 ordered nanocrystals.}\label{fig:3damp_strain}
\end{figure}

Figures \ref{fig:3damp_strain}a-c show the exteriors of the nanocrystals and cross sections of the amplitude fields from the BCDI reconstructions. Figure \ref{fig:3damp_strain}a shows the three disordered nanocrystals, Fig. \ref{fig:3damp_strain}b shows the semi-ordered nanocrystals, and Fig. \ref{fig:3damp_strain}c shows the ordered nanocrystals. For all nanocrystals shown, an amplitude threshold of $5\times10^{-5}$ is used to define a nanocrystal. The amplitude fields are correlated to the local average electron density (or atomic number) in addition to the local ordering at each position. In the amplitude fields, regions of higher and lower scattering power can be observed through the cross sections of all the specimens. The disordered and semi-ordered nanocrystals exhibit a mix of both lower frequency and higher frequency fluctuations of the scattering amplitude. The ordered specimens have significantly fewer higher frequency fluctuations, and instead, all have regions of significantly higher scattering towards the center of the nanocrystals, producing a visible `core-shell' structure.

Similarly, Figures \ref{fig:3damp_strain}d-f show the reconstructed exteriors and cross sections of the 111 lattice strain fields $\varepsilon_{111}$ calculated from the phase $\phi$ of the scattering reconstructions. Note that the largest strains are truncated (magnitude greater than 0.002) in order to increase visibility, but the full histograms are provided below. As with the amplitude fields, higher frequency fluctuations are observed in the disordered (Fig. \ref{fig:3damp_strain}d) and semi-ordered (Fig. \ref{fig:3damp_strain}e) nanocrystals. The maximum and minimum strain magnitudes also are generally larger in the disordered and semi-ordered nanocrystals than in the ordered nanocrystals. We note that in other work \cite{moniri2023three}, the nanoscale strain fields in a high entropy alloy were found to be of higher frequency and magnitude than in a corresponding medium entropy alloy, where presumably the high entropy alloy is more disordered. Here, there also appears to be correlations between locations of fluctuating amplitude and the regions where strain in the Cu$_3$Au is of high magnitude which will be further examined.

\begin{figure}[h!]
    \centering
    \includegraphics[width = 1.0 \textwidth]{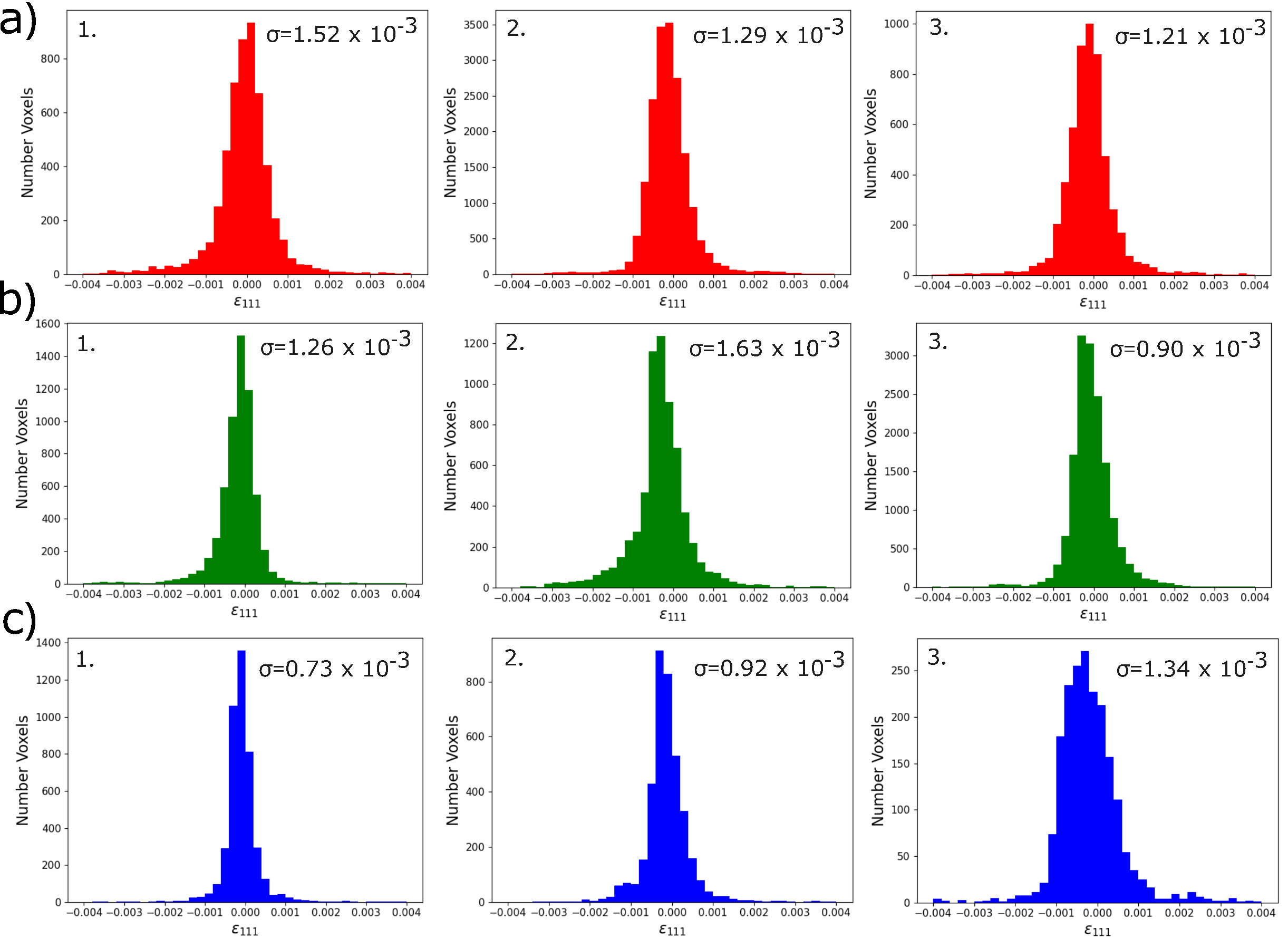}
    \caption{Histograms of the voxel 111 lattice strains $\varepsilon_{111}$ of the a) disordered (red), b) semi-ordered (green), and c) ordered crystals (blue). Standard deviations $\sigma$ of the lattice strain distributions also provided.}\label{fig:histograms}
\end{figure}

Figures \ref{fig:histograms}a-c show the full lattice strain distributions of the disordered (red), semi-ordered (green), and ordered (blue) specimens in histogram form. Also reported are the standard deviations $\sigma$ of the distributions. On average, the distributions of strains of the disordered (mean $\sigma=1.34\times10^{-3}$) and semi-ordered (mean $\sigma=1.26\times10^{-3}$) nanocrystals are broader than the ordered nanocrystals (mean $\sigma=0.99\times10^{-3}$). We interpret the smaller distribution widths of the ordered nanocrystals arising from local (voxel) lattice parameters being nearer to the nanocrystal mean. However, while there is a trend, there is also significant scatter. Future studies with larger numbers of nanocrystals would help improve the statistics and more concretely build a relationship between spatial lattice strain distributions and ordering.

\begin{figure}[h!]
    \centering
    \includegraphics[width = 0.7 \textwidth]{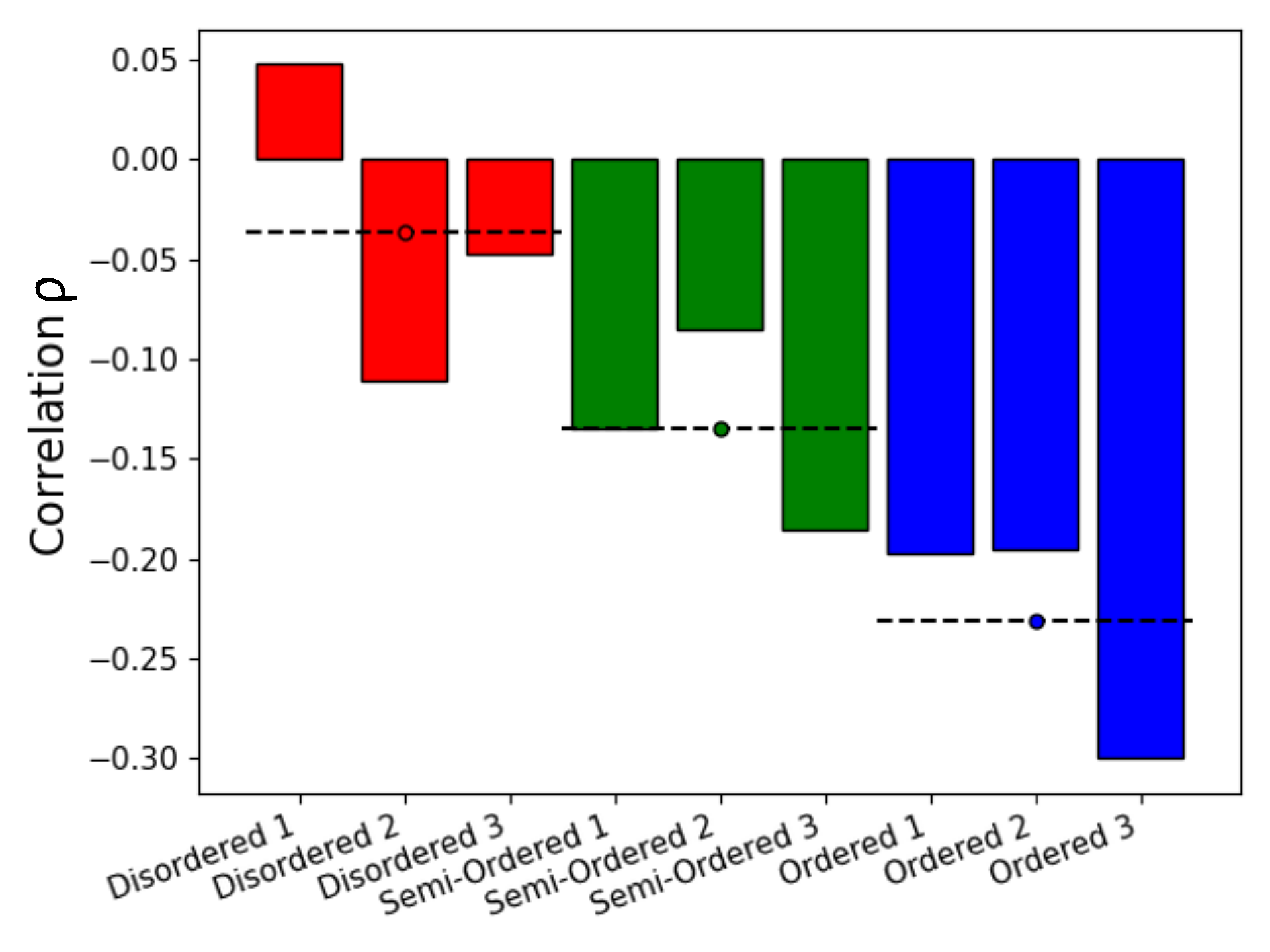}
    \caption{Correlation metric $\rho$ (Pearson correlation coefficient) between scattering amplitude $A$ and 111 lattice strain $\varepsilon_{111}$ for all nanocrystals. The mean correlation for each heat treatment is marked with a dashed line.}\label{fig:corr}
\end{figure}

To examine spatial correlations between the scattering amplitude and lattice strain fields, a correlation coefficient $\rho$ was calculated for each nanocrystal. We used a Pearson correlation coefficient $\rho$ defined as
\begin{equation}
    \rho=\frac{\bar{\varepsilon}_{111}(\bm{r}) \cdot \bar{A}(\bm{r})}{\sqrt{\bar{\varepsilon}_{111}(\bm{r})\cdot \bar{\varepsilon}_{111}(\bm{r})}\sqrt{\bar{A}(\bm{r}) \cdot \bar{A}(\bm{r})}}
\end{equation}
where the dot product is performed over all voxels and the bar indicates that quantities have been mean-subtracted. This value $\rho$ will range from -1 to 1, with higher magnitudes indicating increased correlation, while a value of 0 indicates no correlation. Note that for this metric, a negative value corresponds to large voxel scattering amplitudes residing at positions with low/negative strains, or the converse. Figure \ref{fig:corr} shows the correlations $\rho$ calculated for all 9 nanocrystals with disordered crystals (colored red), semi-ordered (colored green), and ordered (colored blue). The mean correlation for the three heat treatments are also indicated (dashed lines). Across the three heat treatments, there is a trend for increased correlation as ordering is increased.

The negative correlation between amplitude and strain (for all nanocrystals except Disordered 1) points to the increases in local scattering being attributed to increased Cu$_3$Au ordering (reduced strain) rather than elemental segregation.  As the chemical ordering of Cu$_3$Au increases, the constructive interference at the 111 Bragg peak also increases. Conversely as ordering drops, more intensity is transferred to diffuse scattering between Bragg peaks \cite{cowley1950approximate}. If the increase in scattering was due to generally higher concentrations of Au (larger electron density) as a result of elemental segregation, we would expect a positive correlation as Au has a larger atomic radius than Cu which would result in an effective increase in lattice parameter and, thus, an increased strain at the boundary of the local Au region. In contrast, here we observe negative correlations between scattering amplitude and lattice strain, indicating that the higher scattering amplitudes can be attributed to increased chemical ordering. This interpretation also implies that ordered regions (higher scattering amplitude) tend to be under compression, while more disordered regions (lower scattering amplitude) are in tension. Qualitatively, this can be seen by comparing the cross-sectional views of the amplitude and strain for any given particle in Figure \ref{fig:3damp_strain}. With regards to the repeatedly appearing core-shell structure in the ordered crystals, it is unclear if this is a surface effect or inherent to the ordering process. A further study examining crystals of different sizes would help to elucidate this phenomenon. Importantly, these increased correlations provide a metric for non-destructively evaluating levels of ordering in alloy systems, necessary for building connections with other properties.

With the completed upgrades of diffraction-limited storage rings \cite{chenevier2018esrf}, the ability to perform BCDI at higher rates and higher resolution for \emph{in situ} microstructural evolution studies, with more coherent beams for the reconstruction of larger crystals, and at higher X-ray energies to probe crystals/grains embedded within polycrystals is now feasible \cite{maddali2019phase,maddali2020high}. Here we have demonstrated that BCDI can be used to non-destructively probe ordering at the nanoscale with 3D insight and without many of the data analysis complications encountered using traditional scattering methods. In the future, BCDI measurements could aid the optimization of SRO for enhanced mechanical performance when applied to studying bulk metallic alloys. In total, we found that in the model Cu$_3$Au system:
\begin{itemize}
    \item There is decreased high frequency fluctuations of scattering amplitude and lattice strain with increased ordering.
    \item Increased presence of regions of localized higher strain magnitudes are found with decreased ordering.
    \item An inverse correlation exists between scattering amplitude and local strain which is interpreted here as regions with increased ordering being in mechanical compression.
\end{itemize}
Further study is required to determine the broader generality of these findings to other alloy systems.

\section*{Acknowledgments:}
Part of this work (experimental measurement and data interpretation) was supported by the U.S. Department of Energy, Office of Science, Basic Energy Sciences, Materials Sciences and Engineering Division.  K.J.H. was supported by Laboratory Directed Research and Development (LDRD) funding from Argonne National Laboratory, provided by the Director, Office of Science, of the U.S. Department of Energy under Contract No. DE-AC02-06CH11357.  This research used resources of the Advanced Photon Source, a US Department of Energy (DOE) Office of Science user facility at Argonne National Laboratory, and is based on research supported by the U.S. DOE Office of Science-Basic Energy Sciences, under Contract No. DE-AC02-06CH11357.

\bibliographystyle{elsarticle-num}
\bibliography{ref.bib}

\begin{thebibliography}{10}
\expandafter\ifx\csname url\endcsname\relax
  \def\url#1{\texttt{#1}}\fi
\expandafter\ifx\csname urlprefix\endcsname\relax\def\urlprefix{URL }\fi
\expandafter\ifx\csname href\endcsname\relax
  \def\href#1#2{#2} \def\path#1{#1}\fi

\bibitem{cohen1962some}
J.~Cohen, M.~Fine, Some aspects of short-range order, J. Phys. Radium 23~(10)
  (1962) 749--762.

\bibitem{rudman1956effect}
P.~S. Rudman, B.~L. Averbach, The effect of cold work on local order, Acta
  Metallurgica 4~(4) (1956) 382--384.

\bibitem{zhang2020short}
R.~Zhang, S.~Zhao, J.~Ding, Y.~Chong, T.~Jia, C.~Ophus, M.~Asta, R.~O. Ritchie,
  A.~M. Minor, Short-range order and its impact on the crconi medium-entropy
  alloy, Nature 581~(7808) (2020) 283--287.

\bibitem{han2024ubiquitous}
Y.~Han, H.~Chen, Y.~Sun, J.~Liu, S.~Wei, B.~Xie, Z.~Zhang, Y.~Zhu, M.~Li,
  J.~Yang, et~al., Ubiquitous short-range order in multi-principal element
  alloys, Nature Communications 15~(1) (2024) 6486.

\bibitem{wu2021short}
Y.~Wu, F.~Zhang, X.~Yuan, H.~Huang, X.~Wen, Y.~Wang, M.~Zhang, H.~Wu, X.~Liu,
  H.~Wang, et~al., Short-range ordering and its effects on mechanical
  properties of high-entropy alloys, Journal of Materials Science \& Technology
  62 (2021) 214--220.

\bibitem{cowley1950approximate}
J.~M. Cowley, An approximate theory of order in alloys, Physical Review 77~(5)
  (1950) 669.

\bibitem{proffen2002chemical}
T.~Proffen, V.~Petkov, S.~Billinge, T.~Vogt, Chemical short range order
  obtained from the atomic pair distribution function, Zeitschrift f{\"u}r
  Kristallographie-Crystalline Materials 217~(2) (2002) 47--50.

\bibitem{owen2017analysis}
L.~Owen, H.~Playford, H.~Stone, M.~Tucker, Analysis of short-range order in
  cu3au using x-ray pair distribution functions, Acta Materialia 125 (2017)
  15--26.

\bibitem{okamoto1971short}
P.~Okamoto, G.~Thomas, On short range order and micro-domains in the ni4mo
  system, Acta Metallurgica 19~(8) (1971) 825--841.

\bibitem{stobbs1978classification}
W.~Stobbs, J.-P. Chevalier, The classification of short range order by electron
  microscopy, Acta Metallurgica 26~(2) (1978) 233--240.

\bibitem{sato2011order}
K.~Sato, A.~Kov{\'a}cs, Y.~Hirotsu, Order--disorder transformation in fe--pd
  alloy nanoparticles studied by in situ transmission electron microscopy, Thin
  Solid Films 519~(10) (2011) 3305--3311.

\bibitem{moniri2023three}
S.~Moniri, Y.~Yang, J.~Ding, Y.~Yuan, J.~Zhou, L.~Yang, F.~Zhu, Y.~Liao,
  Y.~Yao, L.~Hu, et~al., Three-dimensional atomic structure and local chemical
  order of medium-and high-entropy nanoalloys, Nature 624~(7992) (2023)
  564--569.

\bibitem{marceau2015quantitative}
R.~K. Marceau, A.~V. Ceguerra, A.~J. Breen, D.~Raabe, S.~P. Ringer,
  Quantitative chemical-structure evaluation using atom probe tomography:
  Short-range order analysis of fe--al, Ultramicroscopy 157 (2015) 12--20.

\bibitem{he2024quantifying}
M.~He, W.~J. Davids, A.~J. Breen, S.~P. Ringer, Quantifying short-range order
  using atom probe tomography, Nature Materials 23~(9) (2024) 1200--1207.

\bibitem{robinson2001reconstruction}
I.~K. Robinson, I.~A. Vartanyants, G.~Williams, M.~Pfeifer, J.~Pitney,
  Reconstruction of the shapes of gold nanocrystals using coherent x-ray
  diffraction, Physical review letters 87~(19) (2001) 195505.

\bibitem{pfeifer2006three}
M.~A. Pfeifer, G.~J. Williams, I.~A. Vartanyants, R.~Harder, I.~K. Robinson,
  Three-dimensional mapping of a deformation field inside a nanocrystal, Nature
  442~(7098) (2006) 63--66.

\bibitem{robinson2009coherent}
I.~Robinson, R.~Harder, Coherent x-ray diffraction imaging of strain at the
  nanoscale, Nature materials 8~(4) (2009) 291--298.

\bibitem{vaxelaire2014new}
N.~Vaxelaire, S.~Labat, T.~W. Cornelius, C.~Kirchlechner, J.~Keckes,
  T.~Schulli, O.~Thomas, New insights into single-grain mechanical behavior
  from temperature-dependent 3-d coherent x-ray diffraction, Acta materialia 78
  (2014) 46--55.

\bibitem{cowley1950x}
J.~M. Cowley, X-ray measurement of order in single crystals of cu3au, Journal
  of Applied Physics 21~(1) (1950) 24--30.

\bibitem{kear1964dislocation}
B.~Kear, Dislocation configurations and work hardening in cu3au crystals, Acta
  Metallurgica 12~(5) (1964) 555--569.

\bibitem{maddali2020general}
S.~Maddali, P.~Li, A.~Pateras, D.~Timbie, N.~Delegan, A.~Crook, H.~Lee,
  I.~Calvo-Almazan, D.~Sheyfer, W.~Cha, et~al., General approaches for
  shear-correcting coordinate transformations in bragg coherent diffraction
  imaging. part i, Journal of Applied Crystallography 53~(2) (2020) 393--403.

\bibitem{maddali2023concurrent}
S.~Maddali, T.~D. Frazer, N.~Delegan, K.~J. Harmon, S.~E. Sullivan, M.~Allain,
  W.~Cha, A.~Dibos, I.~Poudyal, S.~Kandel, Y.~S.~G. Nashed, F.~J. Heremans,
  H.~You, Y.~Cao, S.~O. Hruszkewycz, Concurrent multi-peak bragg coherent x-ray
  diffraction imaging of 3d nanocrystal lattice displacement via global
  optimization, npj Computational Materials 9~(1) (2023) 77.

\bibitem{hofmann2018glancing}
F.~Hofmann, R.~J. Harder, W.~Liu, Y.~Liu, I.~K. Robinson, Y.~Zayachuk,
  Glancing-incidence focussed ion beam milling: A coherent x-ray diffraction
  study of 3d nano-scale lattice strains and crystal defects, Acta Materialia
  154 (2018) 113--123.

\bibitem{hofmann2020nanoscale}
F.~Hofmann, N.~W. Phillips, S.~Das, P.~Karamched, G.~M. Hughes, J.~O. Douglas,
  W.~Cha, W.~Liu, Nanoscale imaging of the full strain tensor of specific
  dislocations extracted from a bulk sample, Physical Review Materials 4~(1)
  (2020) 013801.

\bibitem{fienup1982phase}
J.~R. Fienup, Phase retrieval algorithms: a comparison, Applied optics 21~(15)
  (1982) 2758--2769.

\bibitem{chenevier2018esrf}
D.~Chenevier, A.~Joly, Esrf: inside the extremely brilliant source upgrade,
  Synchrotron Radiation News 31~(1) (2018) 32--35.

\bibitem{maddali2019phase}
S.~Maddali, M.~Allain, W.~Cha, R.~Harder, J.-S. Park, P.~Kenesei, J.~Almer,
  Y.~Nashed, S.~O. Hruszkewycz, Phase retrieval for bragg coherent diffraction
  imaging at high x-ray energies, Physical Review A 99~(5) (2019) 053838.

\bibitem{maddali2020high}
S.~Maddali, J.-S. Park, H.~Sharma, S.~Shastri, P.~Kenesei, J.~Almer, R.~Harder,
  M.~J. Highland, Y.~Nashed, S.~O. Hruszkewycz, High-energy coherent x-ray
  diffraction microscopy of polycrystal grains: Steps toward a multiscale
  approach, Physical Review Applied 14~(2) (2020) 024085.

\end{thebibliography}




\end{document}